\documentclass[a4paper,11pt,preprint,showpacs,superscriptaddress,preprintnumbers,prd]{revtex4}
\usepackage{graphicx,amsmath,dcolumn,bm}
\usepackage{hyperref}
\usepackage{epstopdf}
\begin{document}
\title{\Large{\bf{Atomic mass, Bjorken variable and scale dependence of quark transport coefficient in Drell-Yan process for proton incident on nucleus}}}

\author{Wei-Jie Xu}
\email[E-mail:]{wjxu@hebtu.edu.cn}
\affiliation{College of Physics and Hebei Advanced Thin Film Laboratory, Hebei Normal University, Shijiazhuang 050024, P.R.China}

\author{Tian-Xing Bai}
\email[E-mail:]{txbai@hebtu.edu.cn}
\affiliation{College of Physics and Hebei Advanced Thin Film Laboratory, Hebei Normal University, Shijiazhuang 050024, P.R.China}

\author{Chun-Gui Duan}
\email[E-mail:]{duancg@hebtu.edu.cn}
\affiliation{College of Physics and Hebei Advanced Thin Film Laboratory, Hebei Normal University, Shijiazhuang 050024, P.R.China}

\begin{abstract}

By means of the nuclear parton distributions determined without the fixed-target Drell-Yan
experimental data and the analytic expression of quenching weight based on BDMPS formalism,
a next-to-leading order analyses are performed on the Drell-Yan differential cross section ratios from Fermilab E906 and E866 Collaborations.
It is found that the calculated results with only the nuclear effects of parton distribution are not in agreement with the E866 and E906 experimental data.
The incoming parton energy loss effect can not be ignored in the nuclear Drell-Yan reactions.
The predicted results indicate that with the quark transport coefficient as a constant,
the suppression due to the target nuclear geometry effect is approximately $19.24\%$ for the quark transport coefficient.
It is shown that we should consider the target nuclear geometry effect in studying the Drell-Yan reaction on nuclear targets.
On the basis of Bjorken variable and scale dependence of the quark transport coefficient, the atomic mass dependence is incorporated.
The quark transport coefficient is determined as a function of the atomic mass, Bjorken variable $x_2$ and scale $Q^2$ by the global fit of the experimental data.
The determined constant factor $\hat{q}_0$ of the quark transport coefficient is $0.061\pm0.004$ GeV$^2$/fm.
It is found that the atomic mass dependence has a remarkable impact on the constant factor $\hat{q}_0$ in the quark transport coefficient in cold nuclear matter.

\vskip 0.1cm



\end{abstract}

\maketitle

\newpage
\vskip 0.5cm

\section{Introduction}

The insight into quark-gluon plasma(QGP) properties has been one of the most active frontiers in nuclear physics and particle physics up to now.
Jet quenching can provide a unique window into the nature of the QGP produced in heavy-ion collisions[1,2].
In order to realize jet quenching, it is necessary to make certain of the energy loss mechanism theoretically because
the initial scattering partons with high transverse momentum propagate through the QGP while losing its energy.
However, the energy loss mechanism is still not clearly known[3-9].

As for the Drell-Yan reaction[10] of hadron on nucleus and the semi-inclusive deep inelastic
scattering of lepton on nucleus, the bound nucleons in target nucleus play the
role of very nearby detectors for the traversing partons.
The hard scattering is localized in space. The properties of target nuclei are well known.
In addition, the semi-inclusive deep inelastic scattering on nuclear targets is
an ideal tool to study the energy loss of the outgoing parton in the cold nuclear medium[11-15].
The hadron-induced Drell-Yan reaction on nuclear targets is an excellent process
to investigate the incoming parton energy loss in cold nuclear matter[16-23].
Therefore, the Drell-Yan reaction of hadron on nucleus and the semi-inclusive deep inelastic
scattering of lepton on nucleus can provide the essential information on the energy loss mechanism of fast partons
in cold nuclear matter which help to explore the similar process appearing in relativistic heavy ion collisions.

According to the parton model interpretation, the Drell-Yan process in hadron-nucleus collisions is closely
related to the parton distribution functions in nuclei.
It is well known that nuclear parton distribution functions are mutually different from those in free nucleons[24].
The nuclear modifications to the nucleon parton distribution functions
are usually referred to as the nuclear effects on the parton distribution functions.
In consideration of the strong necessity of precise nuclear parton distributions, the global analysis of nuclear parton
distribution functions has been employed for over twenty years[25].
Several groups have presented their global analysis of the nuclear parton distribution functions.
They are Eskola et al.(EKS98[26], EKPS[27], EPS09[28] and EPPS16[29]),
Hirai et al.(HKM[30], HKN04[31], and HKN07[32]),
de Florian et al.(nDS[33] and DSZS[34]), I. Schienbein et al.[35],
K. Kovarik et al.(nCTEQ15[36]), Marina Walt et al.(TUJU19[37]) and
Rabah Abdul Khalek et al.(nNNPDF1.0[38] and nNNPDF2.0[39]).
It is noticeable that HKM, TUJU19, nNNPDF1.0 and nNNPDF2.0 proposed the nuclear parton distributions
that did not employ the existing experimental data on the fixed-target Drell-Yan process.

As for hadron-nucleus Drell-Yan process, the energy loss of incoming partons is another nuclear effect
apart from the nuclear effects on the parton distributions as in deep inelastic scattering.
In order to investigate the energy loss of a high energy parton in the cold nuclear matter,
Baier et al.[40,41] (BDMPS hereafter) treated the multiple scattering of the high energy quark in the nucleus by the Glauber approximation.
The medium-induced transverse momentum broadening and induced gluon radiation spectrum of a high energy quark traversing a nucleus were studied.
BDMPS showed the radiative energy loss of the parton per unit length, grows as the length of the nuclear matter, as does the transverse momentum broadening.
In the BDMPS formalism, the quark transport coefficient not only determines the energy loss,
but it is also related to transverse momentum broadening of energetic partons propagating in the medium.
The quark transport coefficient is defined as
\begin{equation}
\hat{q}=\frac{4\pi^2\alpha_s(Q^2_{\rm G})C_F}{N^2_c-1}\rho x_{\rm G} G(x_{\rm G},Q^2_{\rm G}),
\end{equation}
where $C_F$ is the quark colour factor, the number of colors $N_c=3$, $\rho$ is the nuclear matter density.
The strong coupling constant $\alpha_s$ and the gluon distribution function $G(x_{\rm G},Q^2_{\rm G})$ depend on the virtuality $Q^2_{\rm G}$.
The relative calculation indicates that the Bjorken variable $x_{\rm G}\ll1$.
The virtuality $Q^2_{\rm G}$ is equal to transverse momentum broadening of energetic partons.
The experimental data from CERN experiment NA10[42] and Fermilab experiment E772[43] show that the virtuality $Q^2_{\rm G}$ is less than 1 GeV$^2$.
However, it is important to keep in mind that the Bjorken variable $x_{\rm G}$ estimated in theory is an unmeasurable kinematic variable in experiment.
The fact limits the predictive power of BDMPS theory.

In order to improve the predictive power of theory, we explored the relation between quark transport coefficient
and the measurable kinematic variable in deep inelastic scattering[15] or nuclear Drell-Yan reaction[22].
By means of the semi-inclusive deep inelastic scattering of lepton on nucleus,
we used the analytic parameterization of quenching weight based on BDMPS formalism[44] with the target nuclear geometry effect.
The hadron multiplicity ratios were calculated with comparison to the HERMES charged pions production data[45]
on the quarks hadronization occurring outside the nucleus.
The relation is discovered between quark transport coefficient and the measurable kinematic variables in deep inelastic scattering.
The quark transport coefficient is determined as a function of the Bjorken variable and the photon virtuality[15].
By means of the proton-nucleus Drell-Yan process, we used the HKM nuclear parton distributions[30] determined only with lepton-nuclear
deep inelastic scattering experimental data and the analytic parameterization of quenching weight based on BDMPS formalism.
The nuclear Drell-Yan differential cross section ratios were computed as a function of Feynman variable
from Fermilab E906[46] and E866[47] experimental data.
The relation is discovered between quark transport coefficient and the measurable kinematic variables in Drell-Yan process.
The quark transport coefficient is determined as a function of the invariant mass of the lepton pair
and the momentum fraction of the partons in target nucleus[22].
In the two published papers[15,22], our theoretical calculations were performed at
leading order QCD approximation. The atomic mass dependence is not involved of quark transport coefficient.

In fact, a high energy parton encounters the various nucleons in the target nucleus when it traverses cold nuclear matter.
In the definition of quark transport coefficient,
the gluon distribution function $G(x_{\rm G},Q^{2}_{\rm G})$ should be that of the bound nucleon in the target nucleus.
In other word, the gluon distribution function $G(x_{\rm G},Q^{2}_{\rm G})$ depends on the atomic mass number.
In the present article, the next-to-leading order analysis are performed of the differential cross section
ratios in the nuclear Drell-Yan process from Fermilab E906[46] and E866[47] experimental data.
The nuclear effects on the parton distribution functions, energy loss effect based on BDMPS formalism
and the target nuclear geometry effect are discussed respectively.
The atomic mass, Bjorken variable and scale dependence of quark transport coefficient is explored for the first time.
It is hoped to gain new knowledge about the quark transport coefficient in cold nuclear medium.

The remainder of the paper is organized as follows.
In Section II, the briefly formalism for the differential cross section in the nuclear Drell-Yan process is presented.
In Section III, the results and discussion obtained are presented.
Finally, a summary is presented.

\section{Brief formalism for differential cross section in nuclear Drell-Yan reaction}

As for proton-nucleus Drell-Yan process[10], the perturbative QCD leading order contribution
is quark-antiquark annihilation into a lepton pair of mass $M$.
In the collinear factorization approach, the leading order(LO) Drell-Yan cross section
in Feynman variable $x_{\rm F}$ and $M$ is given by
\begin{equation}
\frac{d^2\sigma^{\rm LO}}{dMdx_{\rm F}}=\frac{8\pi\alpha_{\rm em}^2}{9Ms}\frac{1}{x_1+x_2}H^{\rm LO}(x_1,x_2,Q^2),
\end{equation}
with
\begin{equation}
H^{\rm LO}(x_1,x_2,Q^2)=\sum_{f}e^2_f[q^{\rm p}_f(x_1,Q^2)\bar{q}^{\rm A}_f(x_2,Q^2)+\bar{q}^{\rm p}_f(x_1,Q^2)q^{\rm A}_f(x_2,Q^2)],
\end{equation}
where $\alpha_{\rm em}$ is the fine structure coupling constant, $\sqrt{s}$ is the center of mass energy of the hadronic collision,
$x_1 $(respectively $x_2$) is the momentum fraction carried by the projectile (respectively target) parton,
the factorization scale $Q^2=M^2$, the sum is carried out over the light flavor $f={\rm u,d,s}$,
$q^{\rm p(A)}_{f}(x,Q^2)$ and ${\bar q}^{\rm p(A)}_{f}(x,Q^2)$ are the quark and antiquark distributions in the proton (nucleon in the nucleus A).

In the perturbative QCD next-to-leading order(NLO) approximation,
additional emission of a parton (quark or gluon) into the final state has to be taken into account.
The next-to-leading order contribution includes
quark-antiquark annihilation processes ($q+\bar{q}\rightarrow\gamma^*+g$) and gluon Compton scattering ($q+g\rightarrow\gamma^*+q$).
The NLO correction to the Drell-Yan cross section is given by
\begin{equation}
\frac{d^2\sigma^{\rm NLO}}{dMdx_{\rm F}}=\frac{8\pi\alpha_{\rm em}^2}{9Ms}\frac{\alpha_s(M^2)}{2\pi}\int^1_0 dz \frac{1}{x_1+x_2}H^{\rm NLO}(x_1,x_2,Q^2,z),
\end{equation}
with
\begin{eqnarray}
H^{\rm NLO}(x_1,x_2,Q^2,z)&=&\sum_{f}e^2_f \{q^{\rm p}_f(x_1,Q^2)\bar{q}^{\rm A}_f(x_2,Q^2)f_q(z)     \nonumber \\
                          &+& g^{\rm p}(x_1,Q^2)[q^{\rm A}_f(x_2,Q^2)+\bar{q}^{\rm A}_f(x_2,Q^2)]f_g(z)+(x_1\leftrightarrow x_2)\},
\end{eqnarray}
where $g^{\rm p(A)}(x,Q^2)$ are the gluon distributions in the proton (nucleon in the nucleus A).
By combining the dimensionless variable $\tau=zx_1x_2$ and the Feynman variable $x_{\rm F}=x_1-x_2$, one can easily find
\begin{equation}
x_1=\frac{1}{2}(\sqrt{x_{\rm F}^2+4(\tau/z)}+x_{\rm F}),\quad x_2=\frac{1}{2}(\sqrt{x_{\rm F}^2+4(\tau/z)}-x_{\rm F}).
\end{equation}
The coefficient functions $f_{q,g}(z)$[48-50] are, in the DIS factorization scheme,
\begin{eqnarray}
f_q(z)&=& C_F[\delta(1-z)(1+\frac{4\pi^2}{3})-6-4z+(\frac{3}{1-z})_{+}+2(1 + z^2)(\frac{\ln(1-z)}{1-z})_{+}],     \nonumber \\
f_g(z)&=& \frac{1}{2}[(z^2+(1-z)^2)\ln(1-z)+\frac{3}{2}-5z+\frac{9}{2}z^2].
\end{eqnarray}
The 'plus' distributions are defined by
\begin{equation}
\int^1_0 dx f(x)[g(x)]_{+}=\int^1_0 dx (f(x)-f(1))g(x).
\end{equation}
Therefore, up to the next to leading order, the differential cross section in a Drell-Yan reaction can be written as
\begin{equation}
\frac{d^2\sigma}{dMdx_{\rm F}}=\frac{d^2\sigma^{\rm LO}}{dMdx_{\rm F}}+\frac{d^2\sigma^{\rm NLO}}{dMdx_{\rm F}}.
\end{equation}

In the hadron-nucleus Drell-Yan reaction, the projectile suffers multiple collisions
and repeated energy losses in the nuclear matter.
In other words, each quark or gluon in the beam hardon can lose a
finite fraction of its energy in the nuclear target due to QCD bremsstrahlung.
After considering the parton energy loss in nuclei,
the incident parton momentum fraction can be shifted from $x_1+\Delta x_1$ to $x_1$ at the point of fusion.
$\Delta x_1 =\Delta E/E_{\rm p}$ with $\Delta E$ ($E_{\rm p}$) the quark energy loss in the nuclear medium (the projectile hadron energy).
After adding the energy loss of the incoming parton in the target nucleus,
\begin{eqnarray}
H^{\rm LO}(x_1,x_2,Q^2) &=& \int_{0}^{(1-x_1)E_{\rm p}}d(\Delta E)P_q(\Delta E,\omega_{c},L)H^{\rm LO}(x_1+\Delta x_1, x_2,Q^2),  \nonumber \\
H^{\rm NLO}(x_1,x_2,Q^2,z)&=& \int_{0}^{(1-x_1)E_{\rm p}}d(\Delta E)\sum_{f}e^2_f \{ P_q(\Delta E,\omega_{c},L)q^{\rm p}_f(x_1+\Delta x_1,Q^2)\bar{q}^{\rm A}_f(x_2,Q^2)f_q(z) \nonumber \\
                      &+& P_g(\Delta E,\omega_{c},L)g^{\rm p}(x_1+\Delta x_1,Q^2)[q^{\rm A}_f(x_2,Q^2)+\bar{q}^{\rm A}_f(x_2,Q^2)]f_g(z)    \nonumber \\
                      &+&(x_1\leftrightarrow x_2)\},
\end{eqnarray}
where the quenching weight $P_{q,g}(\Delta E,\omega_{c},L)$ are the probability that the radiated gluons carry altogether a given energy $\Delta E$
for an incident quark and gluon, respectively.
$\omega_{c}$ is the characteristic gluon frequency, which is equal to $(1/2)\hat{q}L^{2}$ with the path length $L$ traversed by the incoming parton.

If we do not consider the target nuclear geometry effect, the average path length of the incident parton in the nucleus is given by
$L=(3/4)R_A$ for the case of a hard-sphere nucleus.
The nuclear radius $R_A=1.12{A^{1/3}}$ fm with atomic mass number $A$[51].
With adding the target nuclear geometry effect[14], the colored incoming parton interacting at $y$ the coordinate along the direction of the incident quark will traverse the path length $L=\sqrt{R_{A}^{2}-b^{2}}+y$ with $\vec{b}$ its impact parameter.
After taking both nuclear geometry and energy loss effects into account,
\begin{eqnarray}
H^{\rm LO}(x_1,x_2,Q^2) &=& \int d^{2}bdy\rho_{A}(\vec{b},y)\int_{0}^{(1-x_1)E_{\rm p}}d(\Delta E) P_q(\Delta E,\omega_{c},L)H^{\rm LO}(x_1+\Delta x_1, x_2,Q^2),     \nonumber \\
H^{\rm NLO}(x_1,x_2,Q^2,z)&=& \int d^{2}bdy\rho_{A}(\vec{b},y)\int_{0}^{(1-x_1)E_{\rm p}}d(\Delta E)\sum_{f}e^2_f \{ P_q(\Delta E,\omega_{c},L)     \nonumber \\
                       &\times& q^{\rm p}_f(x_1+\Delta x_1,Q^2)\bar{q}^{\rm A}_f(x_2,Q^2)f_q(z)     \nonumber \\
                      &+& P_g(\Delta E,\omega_{c},L)g^{\rm p}(x_1+\Delta x_1,Q^2)[q^{\rm A}_f(x_2,Q^2)+\bar{q}^{\rm A}_f(x_2,Q^2)]f_g(z)   \nonumber \\
                      &+&(x_1\leftrightarrow x_2)\},
\end{eqnarray}
where $\rho_A(\sqrt{b^2+y^2})=(\rho_0/A)\Theta(R_A-\sqrt{b^2+y^2})$ with $\rho_0$ the nuclear density.

\section{Results and discussion}

In the present analysis, the experimental data are taken from E906[46] and E866[47] Collaborations at Fermilab.
E866 Collaboration reported the observation of the ratios of the Drell-Yan cross
section per nucleon for an 800 GeV proton beam incident on Be, Fe and W targets.
The Drell-Yan events were recorded in the range
$4.0<M<8.4$ GeV, $0.01<x_2<0.12$, $0.21<x_1<0.95$ and $0.13<x_{\rm F}<0.93$.
E906 Collaboration performed the observation of the ratios of the Drell-Yan cross
section per nucleon for an 120 GeV proton beam incident on C, Fe and W targets.
Muon pairs were recorded in the range in the range
$4.5< M < 5.5$ GeV and $0.1< x_2 < 0.3$.
To be emphasized, E866 and E906 data cover the momentum fraction of the target parton
from 0.01 to 0.12, and from 0.1 to 0.3, respectively.
The fact makes E866 and E906 data show possibly the measurable kinematic
variables dependence of the quark transport coefficient in the cold nuclear medium.

In order to study the property of the quark transport coefficient,
we calculate the Drell-Yan cross section ratio of two different nuclear targets bombarded by proton,
\begin{equation}
R_{A_1/A_2}(x_{\rm F})=\int\frac{d^2\sigma^{{\rm p}A_1}}{dMdx_{\rm F}}dM\Bigg/\int\frac{d^2\sigma^{{\rm p}A_2}}{dMdx_{\rm F}}dM,
\end{equation}
in perturbative QCD next-to-leading order approximation. The comparison is performed with selected experimental data.
The integral range in above equation is given by means of the relative experimental kinematic region.

In order to research the nuclear effects from parton distribution functions on the Drell-Yan differential cross section ratio,
$\chi^2$ is calculated with the Drell-Yan differential cross section rations $R_{A_1/A_2}$ as
\begin{equation}
\chi^2=\sum_j\frac{(R^{\rm data}_{A_1/A_2 ,j} -R^{\rm theo}_{A_1/A_2 ,j})^2}{(R^{\rm err}_{A_1/A_2 ,j})^2},
\end{equation}
where the experimental error is given by $R^{\rm err}_{A_1/A_2 ,j}$, and $R^{\rm data}_{A_1/A_2 ,j} (R^{\rm theo}_{A_1/A_2 ,j})$ indicates the experimental
data (theoretical) value for the Drell-Yan differential cross section ratio $R_{A_1/A_2}$ as a function of Feynman variable.
We employ the nNNPDF2.0 nuclear parton distribution functions
from the NNPDF collaboration[39] in our calculations without the consideration of quark energy loss effect.
The calculated values of $\chi^2$ per number of degrees of freedom are
$0.432$, $3.174 $, $6.626$ and $7.819$
for the Drell-Yan differential cross section ratio
Fe/Be, W/Be, Fe/C and W/C, respectively.
Our obtained values of $\chi^2/{\rm ndf}$ are 1.803 and 7.222 for the E866 and E906 experimental data, respectively.
In total, the value of $\chi^2/{\rm ndf}$ is given by 4.126 for the E906 and E866 experimental data(see TABLE I).
The theoretical results in the global fit are compared with the measured Drell-Yan differential cross-section ratios from E906 and E866 collaboration in Fig.1(black solid curves).
Obviously, the calculated results with only considering the nuclear effects of parton distribution do not agree with the E866 and E906 experimental data.
In addition, the main contribution of $\chi^2/{\rm ndf}$ is from the E906 experimental data in the global fit.
It is found that the value of $\chi^2/{\rm ndf}$ by the cross-section ratio Fe/Be is remarkably
lower than those from W/Be, Fe/C and W/C.
It is shown that the theoretical prediction is in good agreement with the experimental data on the cross section ratio Fe/Be.
We think that the theoretical result overestimates the nuclear effects on parton distribution functions in the Drell-Yan differential cross section ratio Fe/Be.
Therefore, it is necessary to refine the nuclear parton distribution in the future.
The accurate nuclear parton distribution can highlight the potential of the fixed-target Drell-Yan measurements
to probe the parton energy loss mechanism in a robust manner.


\begin{table}
\caption{The values of $\hat{q}_0$ and $\chi^2/{\rm ndf}$ extracted from
fitting the selected experimental data on the Drell-Yan differential cross section ratio from the E906 and E866 experiments
by considering the nuclear effects on parton distribution functions(NEPDF) and adding energy loss(EL) effect without and with the nuclear geometry effect.}
\begin{ruledtabular}
\begin{tabular*}{\hsize}
{c@{\extracolsep{0ptplus1fil}} c@{\extracolsep{0ptplus1fil}}
c@{\extracolsep{0ptplus1fil}} c@{\extracolsep{0ptplus1fil}}
c@{\extracolsep{0ptplus1fil}}}
          & \multicolumn{2}{c}{E866}& \multicolumn{2}{c}{E906}           \\
          \cline{2-3} \cline{4-5}
           & Fe/Be &W/Be& Fe/C &W/C\\
           \cline{1-5}
                  & $\chi^2/{\rm ndf}=0.432$ & $\chi^2/{\rm ndf}=3.174 $& $\chi^2/{\rm ndf}=6.626$ & $\chi^2/{\rm ndf}=7.819$\\
        \cline{2-3} \cline{4-5}
       NEPDF   & \multicolumn{2}{c}{$\chi^2/{\rm ndf}=1.803$}& \multicolumn{2}{c}{$\chi^2/{\rm ndf}=7.222$}  \\
       \cline{2-5}
                   & \multicolumn{4}{c}{$\chi^2/{\rm ndf}=4.126$}  \\
       \cline{1-5}
                  & $\hat{q}_0=0$ & $\hat{q}_0=1.005\pm0.234 $& $\hat{q}_0=0.718\pm0.061$ & $\hat{q}_0=0.331\pm0.025$\\
        EL & $\chi^2/{\rm ndf}=0.432$ & $\chi^2/{\rm ndf}=0.552 $& $\chi^2/{\rm ndf}=1.567$ & $\chi^2/{\rm ndf}=0.689$\\
        \cline{2-3} \cline{4-5}
       $(L=3/4R_A)$   & \multicolumn{2}{c}{$\hat{q}_0=1.028\pm0.219$}& \multicolumn{2}{c}{$\hat{q}_0=0.346\pm0.022$}  \\
                 & \multicolumn{2}{c}{$\chi^2/{\rm ndf}=0.674$}& \multicolumn{2}{c}{$\chi^2/{\rm ndf}=2.201$}  \\
       \cline{2-5}
                 & \multicolumn{4}{c}{$\hat{q}_0=0.343\pm0.022 \quad\quad \chi^2/{\rm ndf}=1.562$}  \\
       \cline{1-5}
       EL($L=\sqrt{R_{A}^{2}-b^{2}}+y$)  & \multicolumn{4}{c}{$\hat{q}_0=0.277\pm0.001 \quad\quad \chi^2/{\rm ndf}=1.334$}  \\

\end{tabular*}
\end{ruledtabular}
\end{table}

\begin{figure}
\centering
\includegraphics*[width=15cm,height=10.7cm]{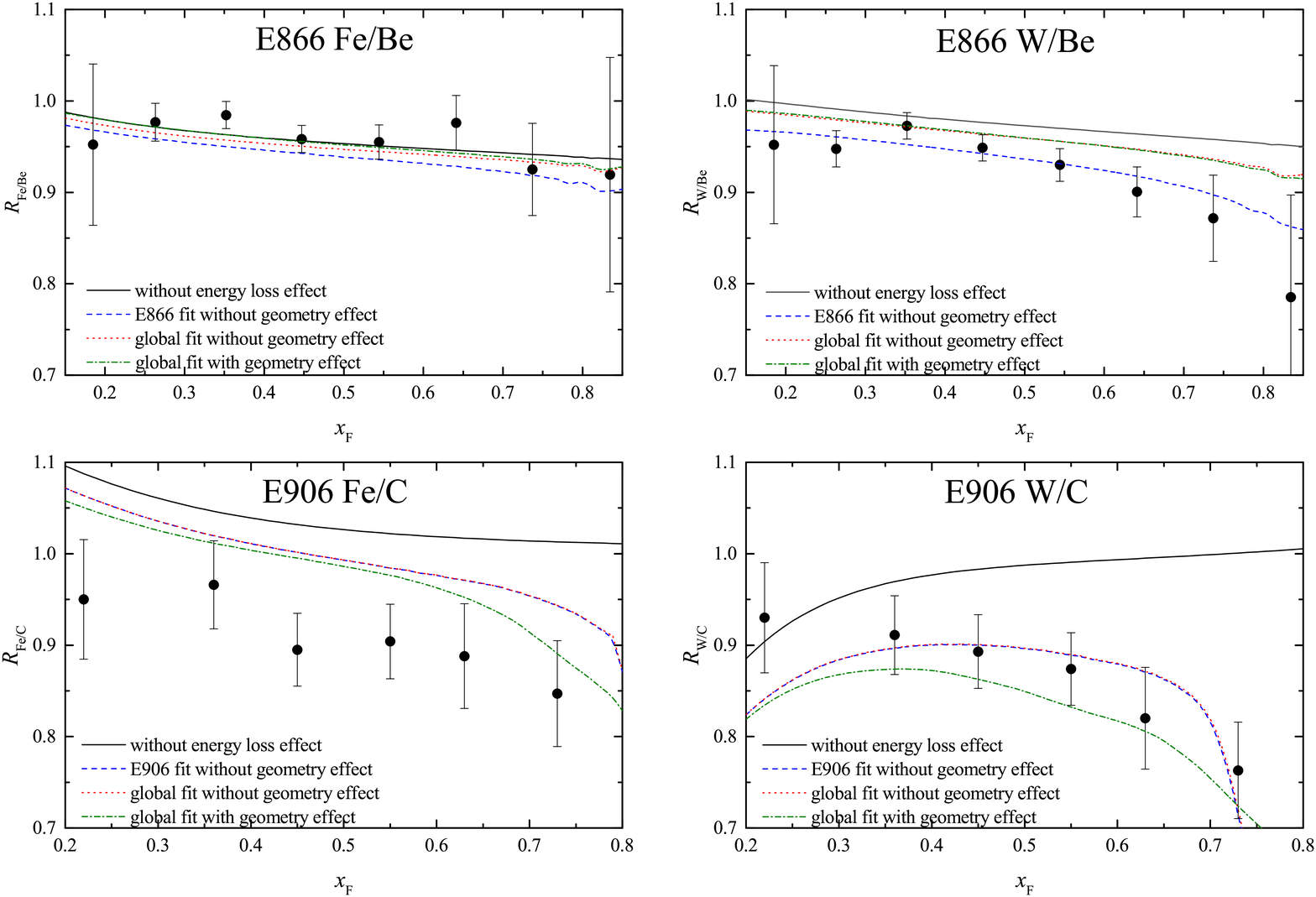}
\vspace{-0.1cm}
\caption{Ratios of the differential cross section per nucleon for Drell-Yan reaction versus Feynman variable $x_{\rm F}$.
The black solid curves are the predicted cross-section ratios in the global fitting E906 and E866 experimental data without energy loss effect.
The blue dashed curves are the predicted results from the fitting respectively E906 and E866 measurements without the target nuclear geometry effect.
As for the global fitting E906 and E866 measurements, the red dotted(green dot-dashed) curves correspond to the theoretical predictions without(with) target nuclear geometry effect.
The experimental data are taken from E906 and E866 Collaboration.}
\end{figure}

On the basis of the nuclear effects on parton distribution functions, the energy loss effect of the incoming parton is added in the following.
The analytic expressions[44] for the quark and gluon quenching weight can be given by using the BDMPS gluon radiation spectrum for an incoming parton in QCD media[52].
The simple analytic expressions can easily be used to compute the Drell-Yan differential cross section.
By means of the CERN subroutine MINUIT[53], the undetermined parameters in the quark transport coefficient $\hat{q}$ can be obtained by minimizing $\chi^2$.

We first make the quark transport coefficient $\hat{q}$ as a constant $\hat{q}_0$ without its kinematic variable dependence.
If we neglect the target nuclear geometry effect, the incoming hard parton will travel an average path length $L=(3/4)R_A$ in the target nuclear medium.
Consequently, only one parameter $\hat{q}_0$ needs to be pinned down.
The values of $\hat{q}_0$ and $\chi^{2}/{\rm ndf}$ extracted from fitting the selected experimental data
on the Drell-Yan differential cross section ratio are listed in TABLE I.
The calculated values of $\hat{q}_0$($\chi^{2}/{\rm ndf}$) are
$0.0$ GeV$^2$/fm $(0.432)$, $1.005\pm0.234 $ GeV$^2$/fm $(0.552)$, $0.718\pm0.061$ GeV$^2$/fm $(1.567)$ and $0.331\pm0.025$ GeV$^2$/fm $(0.689)$
for the Drell-Yan differential cross section ratio Fe/Be, W/Be, Fe/C and W/C, respectively.
The obtained values of $\hat{q}_0$($\chi^{2}/{\rm ndf}$) are
$1.028\pm0.219$ GeV$^2$/fm $(0.674)$ and $0.346\pm0.022 $ GeV$^2$/fm $(2.201)$ for the E866 and E906 experimental data, respectively.
In view of the measured different $x_2$ range in the E906 and E866 experiment,
we can came obviously to the conclusion that the quark transport coefficient $\hat{q}$ depends on the momentum fraction of the target parton.
However, it is a great pity that the global fitting the E906 and E866 experimental data gives
$\hat{q}_0=0.343\pm0.022$ GeV$^2$/fm with $\chi^{2}/{\rm ndf}=1.562$.
Therefore, it is worth intensively looking into the kinematic variable dependence of the quark transport coefficient.
Just to confirm this conclusion, we need more accurate experimental data on the Drell-Yan differential cross section ratio.
We strongly desire that Fermilab Experiment 906 collaboration can refine their experimental data,
and report their precise measurement of the ratios of the Drell-Yan cross section per nucleon in the future published paper.

To demonstrate intuitively the energy loss effect of an incoming parton on the Drell-Yan cross section ratio,
the calculated results combining nNNPDF2.0 nuclear parton distribution functions are compared with the selected experimental data in Fig.1.
The blue dashed curves are the predicted cross-section ratios for the case of fitting respectively E906 and E866 measurements without the target nuclear geometry effect.
As for the global fitting E906 and E866 measurements, the red dotted curves are the theoretical predictions without target nuclear geometry effect.
In particular is worth mentioning that the Drell-Yan cross section ratio of Fe and C nuclear targets bombarded by proton.
As shown in TABLE I and Fig.1,
the agreement is bad between the experimental data and the theoretical results from fitting the E906 measurement and only the cross section ratio Fe/C.
The fact reminds us that the precise measurement of the ratios of the Drell-Yan cross section per nucleon and
accurate nuclear parton distribution functions are indispensable for knowing deeply the energy loss mechanism in the cold nuclear medium.

Furthermore, we explore the energy loss with the target nuclear geometry effect on the Drell-Yan cross section ratio.
In this case, the path length $L=\sqrt{R_{\rm A}^2-b^2}+y$ for the incident parton traversing the nucleus.
Our calculated result indicates that
$\hat{q}_0=0.277\pm0.001$ GeV$^2$/fm with $\chi^{2}/{\rm ndf}=1.334$ from the global fitting the E906 and E866 experimental data.
The comparison of our theorized expectations (green dot-dashed curves) is presented with the experimental measurements in Fig.1.
From TABLE I, it is found that the target nuclear geometry effect could reduce the quark transport coefficient by as much as $19.24\%$.
Therefore, we should consider the target nuclear geometry effect over the course of studying the nuclear Drell-Yan reaction.

Now, let us investigate the kinematic variable dependence of the quark transport coefficient.
In our preceding article[22], we discovered the relation between the Bjorken variable $x_{\rm G}$ in the quark transport coefficient and
the momentum fraction of the target parton in Drell-Yan process.
In other words, the Bjorken variable $x_{\rm G}$ can be replaced with the momentum fraction $x_2$.
Consequently, we can deduce that the virtuality $Q^2_{\rm G}$ should also be expressed with the measurable scale $Q^2$,
which is the square of the lepton pair mass $M$. By means of fitting the experimental data on the Drell-Yan cross section ratio,
the quark transport coefficient has been determined as a function of the Bjorken variable $x_2$ and scale $Q^2$.

As regarding the gluon density $x_{\rm G}G(x_{\rm G},Q^2_{\rm G})$ from the quark transport coefficient $\hat{q}$,
$x_{\rm G}\ll1$, and $Q^2_{\rm G}<1$ GeV$^{2}$[22].
Phenomenological work[54] provided a model based on the concept of saturation for small $Q^2_{\rm G}$ and small $x_{\rm G}$.
A good description of data on the proton structure function was presented by means of the proton saturation scale.
According to the saturation model, the gluon density
\begin{equation}
x_{\rm G}G(x_{\rm G},Q^2_{\rm G})\sim x_{\rm G}^{-\lambda}.
\end{equation}
With using the Bjorken variable $x_2$ instead of $x_{\rm G}$,
\begin{equation}
x_{\rm G}G(x_{\rm G},Q^2_{\rm G})\sim x_2^{-\lambda}.
\end{equation}
Hence, the Bjorken variable $x_2$ dependence of the quark transport coefficient can be written as
\begin{equation}
\hat{q}=\hat{q_0}x_2^{\alpha}.
\end{equation}
Furthermore, adding the intermediate and large $x_2$ correction with the evolution of gluon distribution with $Q^2$,
the quark transport coefficient can be supposed to be
\begin{equation}
\hat{q}(x_2, Q^2)=\hat{q_0} \alpha_{\rm s}(Q^2) x_2^{\alpha}(1-x_2)^{\beta}\ln^{\gamma}(Q^2/Q^2_0),
\end{equation}
where $Q^2_0 = 1$ GeV$^2$ in order to make the argument in the logarithm dimensionless.
Four undetermined parameters are $\hat{q_0}$, $\alpha$, $\beta$ and $\gamma$.

In fact, the gluon density $x_{\rm G}G(x_{\rm G},Q^2_{\rm G})$ in the quark transport coefficient
should be that of the bound nucleon in the target nucleus.
The gluon density $x_{\rm G}G(x_{\rm G},Q^2_{\rm G})$ depends on the atomic mass number in the fixed-target experiment.
The saturation scale of a nucleus[55,56] is enhanced relative to the nucleon one by a factor $A^{1/3}$.
With the atomic mass number $A$ and $x_2$ dependence of the saturation scale in the nucleus, the gluon density
\begin{equation}
x_{\rm G}G(x_{\rm G},Q^2_{\rm G})\sim A^{1/3} x_{\rm G}^{-\lambda} \sim A^{1/3} x_2^{-\lambda}.
\end{equation}
Therefore, we can give the atomic mass, Bjorken variable and scale dependence of the quark transport coefficient,
\begin{equation}
\hat{q}(A, x_2, Q^2)=\hat{q_0} A^{1/3} \alpha_{\rm s}(Q^2) x_2^{\alpha}(1-x_2)^{\beta}\ln^{\gamma}(Q^2/Q^2_0).
\end{equation}
Thus, there are still four parameters to fit: $\hat{q_0}$, $\alpha$, $\beta$ and $\gamma$.

We perform the global analysis of the quark transport coefficient by comparing with the experimental data from the E906 and E866 collaboration.
The parameter values are pinned down in the quark transport coefficient.

\begin{table}
\caption{The parameter values of $\hat{q}$ and $\chi^{2}/{\rm ndf}$ extracted by the global analysis of E866 and E906 experimental data.}
\begin{ruledtabular}
\begin{tabular*}{\hsize}
{c@{\extracolsep{0ptplus1fil}} c@{\extracolsep{0ptplus1fil}}
c@{\extracolsep{0ptplus1fil}} c@{\extracolsep{0ptplus1fil}}
c@{\extracolsep{0ptplus1fil}} c@{\extracolsep{0ptplus1fil}}}

  $\hat{q}$(GeV$^2$/fm)                                            &$\hat{q_0}$       &$\alpha$           &$\beta$           &$\gamma$          &$\chi^{2}/{\rm ndf}$\\
  \hline
  $\hat{q}(x_2, Q^2)$ (without geometry effect)                    &$0.424\pm0.022$   &$-0.083\pm0.026$   &$3.879\pm0.383$   &$1.426\pm0.045$   &$1.561$             \\
  $\hat{q}(x_2, Q^2)$ (with geometry effect)                       &$0.393\pm0.018$   &$-0.106\pm0.006$   &$4.441\pm0.052$   &$1.452\pm0.021$   &$1.218$             \\
  $\hat{q}(A, x_2, Q^2)$ (without geometry effect)                 &$0.076\pm0.005$   &$-0.242\pm0.028$   &$4.399\pm0.452$   &$1.174\pm0.051$   &$1.524$             \\
  $\hat{q}(A, x_2, Q^2)$ (with geometry effect)                    &$0.061\pm0.004$   &$-0.138\pm0.006$   &$4.557\pm0.073$   &$1.448\pm0.033$   &$1.352$             \\

\end{tabular*}
\end{ruledtabular}
\end{table}

\begin{figure}
\centering
\includegraphics*[width=15cm,height=10.7cm]{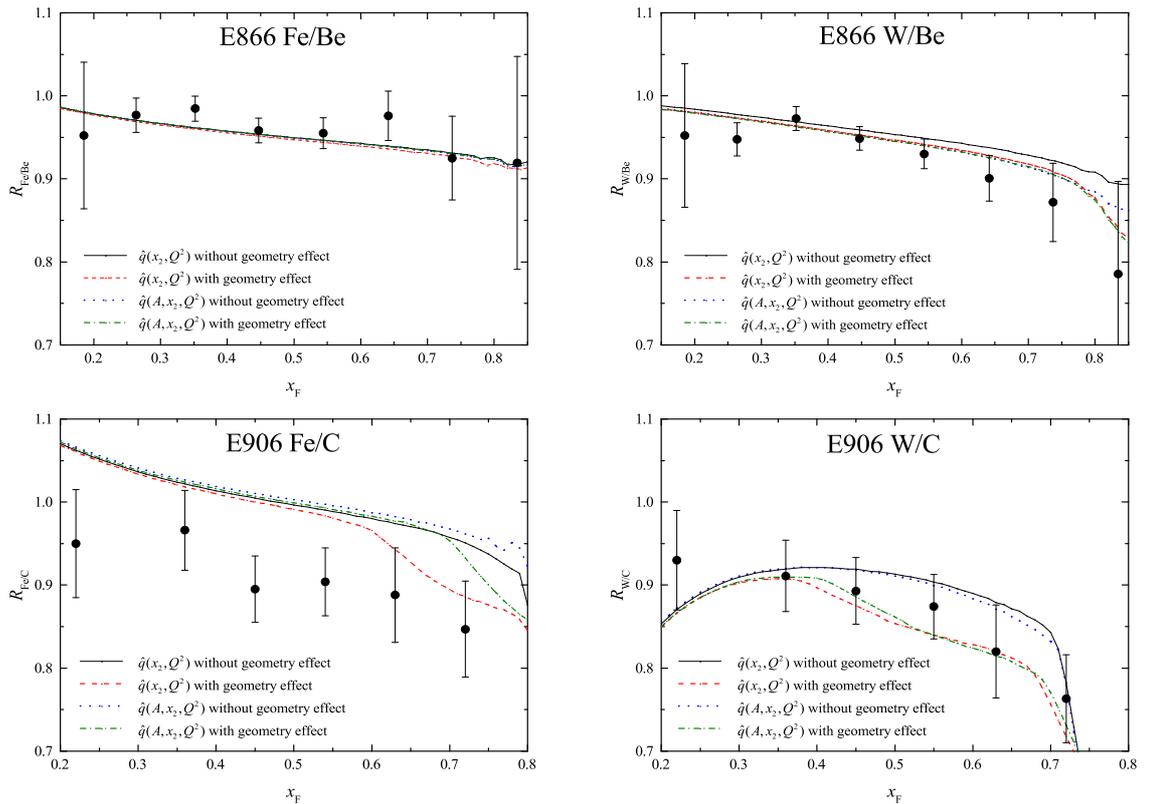}
\vspace{-0.1cm}
\caption{The calculated Drell-Yan cross section ratios $R_{\rm A/B}$ versus Feynman variable $x_{\rm F}$.
The black solid(red dashed) curves are the predicted cross-section ratios from the quark transport coefficient $\hat{q}(x_2, Q^2)$ without(with) target nuclear geometry effect.
The blue dotted(green dot-dashed) curves correspond to the theoretical predictions for the quark transport coefficient $\hat{q}(A, x_2, Q^2)$ without(with) target nuclear geometry effect.
The experimental data are taken from E906 and E866 Collaboration.}
\end{figure}

As for the quark transport coefficient without atomic mass dependence and target nuclear geometry effect, the extracted parameter values are $\hat{q}_0=0.424\pm0.022$ GeV$^2$/fm with the relative uncertainty $\delta \hat{q}_0/\hat{q}_0\simeq5.2\%$, $\alpha=-0.083\pm0.026$ with $\delta\alpha/\alpha\simeq31.3\%$, $\beta=3.879\pm0.383$ with $\delta\beta/\beta\simeq9.9\%$, and $\gamma=1.426\pm0.045$ with $\delta\gamma/\gamma\simeq3.2\%$ and $\chi^2/{\rm ndf}=1.561$.
In regard to the quark transport coefficient with no atomic mass dependence and target nuclear geometry effect, the obtained parameter values are $\hat{q}_0=0.393\pm0.018$ GeV$^2$/fm with  $\delta \hat{q}_0/\hat{q}_0\simeq4.5\%$, $\alpha=-0.106\pm0.006$ with $\delta\alpha/\alpha\simeq5.6\%$, $\beta=4.441\pm0.052$ with $\delta\beta/\beta\simeq1.1\%$, and $\gamma=1.452\pm0.021$ with $\delta\gamma/\gamma\simeq1.4\%$ and $\chi^2/{\rm ndf}=1.218$.
As regards the quark transport coefficient with atomic mass dependence and no target nuclear geometry effect, the theoretical parameter values are $\hat{q}_0=0.076\pm0.005$ GeV$^2$/fm with  $\delta \hat{q}_0/\hat{q}_0\simeq6.6\%$, $\alpha=-0.242\pm0.028$ with $\delta\alpha/\alpha\simeq11.6\%$, $\beta=4.399\pm0.452$ with $\delta\beta/\beta\simeq10.2\%$, and $\gamma=1.174\pm0.051$ with $\delta\gamma/\gamma\simeq4.3\%$ and $\chi^2/{\rm ndf}=1.524$.
In the case of the quark transport coefficient with atomic mass dependence and target nuclear geometry effect, the determined parameter values are $\hat{q}_0=0.061\pm0.004$ GeV$^2$/fm with  $\delta \hat{q}_0/\hat{q}_0\simeq6.5\%$, $\alpha=-0.138\pm0.006$ with $\delta\alpha/\alpha\simeq4.3\%$, $\beta=4.557\pm0.073$ with $\delta\beta/\beta\simeq1.6\%$, and $\gamma=1.448\pm0.033$ with $\delta\gamma/\gamma\simeq2.2\%$ and $\chi^2/{\rm ndf}=1.352$.
The comparison of our theorized results is presented with the experimental measurements in Fig.2.
In Table II, we summarize the extracted parameter values of the quark transport coefficient.
From Table II, it is shown that the atomic mass dependence reduces the constant factor $\hat{q}_0$ in the quark transport coefficient by approximately $82.1\%$ without the target nuclear geometry effect.
After incorporating the target nuclear geometry effect, the atomic mass dependence reduces the constant factor $\hat{q}_0$ in the quark transport coefficient by as much as $84.5\%$.
Therefore, the atomic mass dependence has a significant impact on the constant factor $\hat{q}_0$ in the quark transport coefficient in cold nuclear matter.

Two other research groups have explored the transport coefficient in cold nuclear matter.
Fran\c{c}ois Arleo et al.[21, 57] studied the transport coefficient parameterization as a function of Bjorken variable without the virtuality $Q^2$ evolution.
The coefficient $\hat{q}_0$ is the only parameter of their model. The value of $\hat{q}_0$ is given by $0.07- 0.09$GeV$^2$/fm.
Peng Ru et al.[58] performed the global analysis of the jet transport coefficient in the framework of the generalized QCD factorization formalism.
The jet transport coefficient has the Bjorken-$x$ and scale $Q^2$ dependent parametrization. The coefficient $\hat{q}_0$ is $ 0.0195^{+0.0085}_{-0.0065}$GeV$^2$/fm.
Therefore, it is hoped that the theoretical progress should be made in the future.
With the precision experimental data on the semi-inclusive deep-inelastic scattering and nuclear Drell-Yan reaction,
a quantitative understanding of the parton energy loss mechanism will be thus achieved in cold nuclear matter.

\section{Concluding remarks}

We study the Drell-Yan process in proton-nucleus collisions, supplementing the perturbative QCD factorized formalism with radiative parton energy loss.
By means of the nuclear parton distributions determined without the fixed-target Drell-Yan process experimental data
and the analytic expression of quenching weight based on BDMPS formalism,
a next-to-leading order analyses on the differential cross section ratios from the nuclear Drell-Yan process are performed,
and compared with the experimental data from Fermilab E906 and E866 Collaborations.
It is found that the calculated results with only considering the nuclear effects of parton distribution do not agree with the E866 and E906 experimental data.
The incoming quark energy loss effect is indispensable in nuclear Drell-Yan process.
With the quark transport coefficient as a constant and no target nuclear geometry effect,
the global fitting the E906 and E866 experimental data gives $\hat{q}_0=0.343\pm0.022$ GeV$^2$/fm.
With the target nuclear geometry effect, $\hat{q}_0=0.277\pm0.001$ GeV$^2$/fm.
The suppression due to the target nuclear geometry effect is approximately $19.24\%$ for the quark transport coefficient.
It is shown that we should consider the nuclear geometry effect over the course of studying the Drell-Yan reaction on nuclear targets.
On the basis of Bjorken variable and scale dependence of the quark transport coefficient, the atomic mass dependence is combined.
The quark transport coefficient is determined as a function of the atomic mass, Bjorken variable $x_2$ and scale $Q^2$ by the global fit of the experimental data.
The constant factor $\hat{q}_0$ of the quark transport coefficient are respectively
$0.393\pm0.018$ GeV$^2$/fm and $0.061\pm0.004$ GeV$^2$/fm without and with the atomic mass dependence.
The suppression due to atomic mass dependence is approximately $84.5\%$ for the constant factor $\hat{q}_0$ in the quark transport coefficient.
It is found that the atomic mass dependence has a remarkable effect on the constant factor $\hat{q}_0$ in the quark transport coefficient in cold nuclear matter.

It is well known that there are the quark energy loss effect and the nuclear effects on the parton distributions in the nuclear Drell-Yan process.
The accurate nuclear parton distribution functions are crucial for studying deeply the energy loss mechanism in the cold nuclear medium.
The proton incident nuclei Drell-Yan data can result in an overestimate for nuclear modification of the sea quark distribution function
if leaving the quark energy loss effect out.
Therefore, we suggest that the new global analysis of nuclear parton distribution functions should not employ the available experimental data on the fixed-target hadron-nucleus Drell-Yan process.

The precise measurement of the ratios of the proton-nucleus Drell-Yan cross section per nucleon
is essential for clarifying the energy loss mechanism in the cold nuclear medium.
Our results strongly suggest that Fermilab Experiment 906 collaboration can refine their experimental data,
and report their precise measurement of the ratios of the Drell-Yan cross section per nucleon in the future published paper.

The investigation into the energy loss mechanism in cold nuclear matter helps to understand the similar process appearing in relativistic heavy ion collisions.
The hadron-induced Drell-Yan reaction on nuclei is an excellent process to explore the incoming parton energy loss in cold nuclear matter.
We expect that our determined parametrization of the quark transport coefficient
provides a useful information on the identification of the transport property of the quark-gluon plasma.
It is desirable to operate precise measurements in the nuclear Drell-Yan process from J-PARC experiment[59] and LHCb-SMOG experiment[60].
These new experimental data should shed light on the parton propagation mechanism in nuclear matter.

\vskip 1cm
{\bf Acknowledgments}
This work is supported in part by the National Natural Science Foundation of China(NSFC) under Grants No.11975090 and No.11575052.

\end{document}